\documentclass[prl,twocolumn]{revtex4}
\usepackage{array,amssymb,graphicx}

\def\wideeqn#1#2{
\begin{widetext}
\noindent
\if#1t
\else
    \raisebox{9pt}[0in][0.0in]
    {$\rule{3.4in}{0.4pt}\rule{0.4pt}{6pt}$\hspace{3.6in}}
\fi
#2
\if#1b
\else
    \raisebox{-9pt}[0in][0.0in]
    {\hspace{3.55in}$\rule{0.4pt}{6pt}\rule[6pt]{3.5in}{0.4pt}$}
\fi
\end{widetext}
\noindent
}

\begin{document}

\title{Effective theory of incompressible quantum Hall liquid crystals}

\author{Michael M. Fogler}

\affiliation{Department of Physics, Massachusetts Institute of
Technology, 77 Massachusetts Avenue, Cambridge, Massachusetts 02139}


\begin{abstract}

I propose an effective theory of zero-temperature phases of the quantum
Hall stipes: a smectic phase where the stripes are static and a novel
quantum nematic phase where the positional order is destroyed by
quantum fluctuations. The nematic is viewed as a Bose condensate
of dislocations whose interactions are mediated by a $U(1)$ gauge field.
Collective mode spectrum and the dynamical structure factor in the two
phases are calculated.

\end{abstract}

\maketitle

Stripe phases are extremely common in nature. Once studied mainly in the
context of pattern formation and soft condensed-matter systems
(convection rolls, ferrofluids, diblock co-polymers, {\it
etc.\/}~\cite{Seul_95}), they are now recognized to be important in
venues ranging from neutron stars~\cite{Neutron_star} to neural networks
of the brain~\cite{Chklovskii}. The ``hard'' condensed-matter community
was alerted to the relevance of stripes after their discovery in
transitional metal oxides, especially, high-$T_c$
cuprates~\cite{High_Tc_stripes}. The subject of this Letter is the
stripe phase in another remarkable correlated system: a two-dimensional
(2D) electron liquid in a transverse magnetic field. This particular
phase is formed when the Landau level occupation factor $\nu$ is close
to a half-integer and is larger than some critical value
$\nu_c$~\cite{Fogler_96}. Unlike the still debated case of high-$T_c$,
the theory of the quantum Hall stripes rests on a much more solid
foundation~\cite{Fogler_96,Moessner_96}. The quantity $1 / \nu$ plays
the role of a small parameter controlling the strength of quantum
fluctuations beyond mean-field. Thus, the mean-field stripe solution is
stable and adequately captures the main properties of the ground state
at large $\nu$. Numerical calculations~\cite{Critical_nu,Scarola_01}
indicate that $\nu_c \sim 4$ for the physically relevant case of Coulomb
interaction. This provides a natural explanation of the magnetotransport
anisotropy observed in GaAs heterostructures~\cite{Experiments} at $\nu
\geq \frac92$.

The states at $\nu \sim \nu_c$ are much less understood because the
fluctuations about the mean-field Hartree-Fock solution are large and
several almost degenerate states are in competition. Small changes in
microscopic parameters may therefore radically change the
nature of the ground state and bring to life novel quantum phases.

%
%
\begin{figure}
\includegraphics[width=2.1in,bb=118 308 557 518]{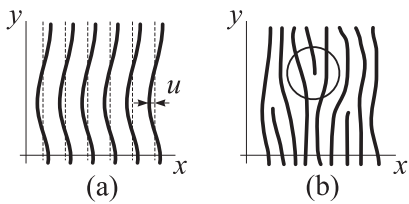}
\setlength{\columnwidth}{3.2in}
\caption{
(a) Smectic (b) Nematic. One dislocation is encircled.
\label{Fig_real_space}
}
\end{figure}

Particularly intriguing are proposed quantum {\it smectic\/} and quantum
{\it nematic\/} phases~\cite{Balents_96,Fradkin_99}. In the smectic
phase the system lacks translational symmetry (is periodically
modulated) along one of the spatial directions, say $\hat{\bf x}$, but
remains liquid-like along $\hat{\bf y}$. An example of the smectic phase
is a unidirectional charge-density-wave Hartree-Fock
state~\cite{Fogler_96}; however, the true quantum smectic must possess
certain amount of fluctuations beyond Hartree-Fock (shown pictorially in
Fig.~\ref{Fig_real_space}a) to prevent another periodic modulation along
$\hat{\bf
y}$~\cite{Fradkin_99,Yi_01,MacDonald_00,Fogler_in_preparation}. In
contrast, the nematic state is fully translationally invariant but lacks
the rotational invariance. It can be visualized as fluctuating stripes
preferentially aligned along the $\hat{\bf y}$-direction, see
Fig.~\ref{Fig_real_space}b. At present, concrete trial wavefunctions for
the quantum nematic are known only for specific filling
fractions~\cite{Musaelian_96}. This supports the idea that this phase is
physically realizable but forces us to seek approaches other than
``wavefunction engineering'' to gain understanding. Below I construct
an effective long-wavelength low-frequency theory that applies to a
class of quantum Hall liquid crystals. It describes zero-temperature
properties and the quantum phase transition between these intriguing
states~\cite{Comment_on_1st_order}.

{\it Smectic phase\/}.--- The smectic emerges from a uniform state when
a pair of collective modes with wavevectors ${\bf q} = \pm q_0 \hat{\bf
x}$ goes soft and condenses~\cite{Comment_on_soft_modes}. Hydrodynamic
fluctuations of the electron density projected onto the topmost Landau
level become of the form $n({\bf r}, t) + {\rm Re}\,\{\Psi({\bf r}, t)
e^{-i q_0 x}\}$, where $n$ is a smooth density component (with $q \ll
q_0$), $\Psi = \Psi_0 e^{i q_0 u}$ is the smectic order parameter, and
$u$ is the deviation of the stripes (density maxima) from uniformity,
see Fig.~\ref{Fig_real_space}a. To build an effective theory I make a
crucial assumption that the Goldstone mode associated with $u$ is {\it
the only\/} low-energy degree of freedom of the system. In general,
other low-energy excitations, e.g., those inherited from the parent
uniform state may be present. Thus, our theory describes only a certain
class of possible smectic states, those that are superstructures on top
of incompressible quantum Hall liquids~\cite{Comment_on_compressible}.
I proceed by writing down the effective Hamiltonian for $u$ and $n$,
containing only the most relevant terms consistent with the
symmetry~\cite{DeGennes_book,Fogler_00}:
\begin{equation}
H = \frac{Y}{2} (\partial_x u)^2
  + \frac{K}{2} (\partial_y^2 u)^2
  + \frac12 \delta n {\bf U} \delta n + C \delta n \partial_x u.
\label{H}
\end{equation}
Here $Y$ and $K$ are phenomenological compression and bending moduli,
$\delta n = n - n_0$ is the deviation of local density from the
equilibrium value $n_0$, and ${\bf U}$ is the integral operator with
kernel $U({\bf r})$, the electron-electron interaction potential. $U$ is
Coulombic at large $r$ but is modified by many-body screening, exchange
and correlation effects at short distances (see more details
in~\cite{Fogler_96}). The last term in Eq.~(\ref{H}) accounts for the
dependence of the smectic period on $n_0$, with $C = Y \partial \ln q_0
/ \partial n_0$. It vanishes at the half-filling due to electron-hole
symmetry but is nonzero otherwise.

One may wonder why include ``incompressible'' background $n$ in our
low-energy Hamiltonian~(\ref{H}). The reason is the important role $n$
plays in the dynamics of $u$. Indeed, $u$ determines the density
fluctuations near the star of the soft mode, e.g., $n_{{\bf q}_0 + {\bf
k}} = \frac12 (\Psi_0 e^{-i q_0 u})_{\bf k}$, while the density
operators $n_{\bf q}$ projected onto a single Landau level are
dynamically linked by the commutation relation~\cite{Girvin_86},
$[n_{\bf q}, n_{{\bf q}^\prime}] = 2 i \sin (\frac12 l^2 {\bf q} \wedge
{\bf q}^\prime) n_{{\bf q} + {\bf q}^\prime}$, where $l = \sqrt{\hbar /
m \omega_c}$ is the magnetic length and $\omega_c = e B / m c$ is the
cyclotron frequency for the external magnetic field $B$. The above
commutation relation is exact but difficult to deal with. I achieve a
simplification by replacing it with its expectation value expanded to
the lowest order in ${\bf q}$. This way I obtain the following
approximate commutation relation for use in our effective theory:
\begin{equation}
 [n_{\bf q}, u_{{\bf q}^\prime}] = (2 \pi)^2 l^2 q_y
\delta({\bf q} + {\bf q}^\prime)
\label{commutator}
\end{equation}
(all other commutators vanish in the $q \to 0$ limit). Introducing the
canonical momentum $p$ by $\delta n = -\partial_y p$, I unify
Eqs.~(\ref{H}) and (\ref{commutator}) into the effective
action~\cite{Comment_on_commutator},
\begin{eqnarray}
 A_{\rm sm} &=& \int\limits_0^\beta d \tau \int d^2 r \left\{
-\frac{i}{l^2} \hbar p \partial_\tau u + \frac{Y}{2} (\partial_x u)^2
  + \frac{K}{2} (\partial_y^2 u)^2
  \right.
\nonumber\\
\mbox{} &+& \left.\frac12 (\partial_y p) {\bf U} (\partial_y p)
  - C \partial_x u \partial_y p
  \right\}.
\label{A_sm}
\end{eqnarray}
This general form can be shown to pass two important tests. First,
after a change of variables it reproduces~\cite{Fogler_in_preparation}
an effective action derived for a specific model of the quantum Hall
smectic~\cite{MacDonald_00}. Second, the density structure factor,
easily calculated for the Gaussian theory~(\ref{A_sm}),
\begin{equation}
S({\bf q}, \omega) \simeq \frac{\hbar}{m}\, {\rm Im}\,
\frac{q_y^2 Q(q)}
{Q(q) \omega_p^2 (q_y / q)^2 - \omega^2 \omega_c^2 - i \omega \delta},
\label{Structure_smectic}
\end{equation}
coincides up to the Bose factor and dissipative terms with the finite
temperature result of Ref.~\cite{Fogler_00}. The notations used here are
$Q({\bf q}) = (\tilde{Y} q_x^2 + K q_y^4) / m n_0$, $\tilde{Y} = Y - C^2
/ U$, and $\omega_p(q) = [n_0 U(q) q^2 / m]^{1/2}$. On the
$\omega > 0$ side, $S({\bf q}, \omega)$ consists of a single
$\delta$-function at the frequency of the magnetophonon mode,
\begin{equation}
\omega({\bf q}) = \frac{q_y}{q}\frac{\omega_p}{\omega_c}
                  \sqrt{Q({\bf q})}.
\label{omega_smectic}
\end{equation}

\begin{figure}
\includegraphics[width=2.1in,bb=122 265 568 513]{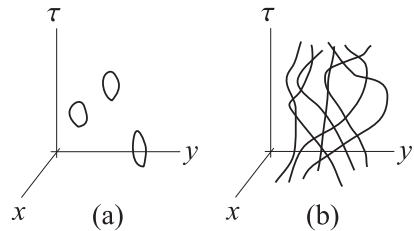}
\vspace{0.1in}
\setlength{\columnwidth}{3.2in}
\caption{
Worldlines of dislocations in (a) smectic (b) nematic.
\label{Fig_worldlines}
}
\end{figure}

{\it Dislocations and duality\/}.--- 2D smectics can exist only at zero
temperature. At $T > 0$ thermal fluctuations of the stripes restore the
translational symmetry, so that the highest possible degree of ordering
is that of the nematic~\cite{DeGennes_book}. The actual crossover of the
Hamiltonian from the short-distance smectic~(\ref{H}) to the
long-distance nematic form is quite nontrivial. It is driven
by thermally excited dislocations, which have an ability to screen the
compressional stress~\cite{Toner_81}. It is natural to assume then that
the smectic-nematic {\it quantum\/} phase transition is also driven by
topological defects. Pictorially, the difference between the smectic and
nematic can be represented as follows. The dislocations are viewed as
lines in the (2 + 1)D space. In the smectic phase, they form small
closed loops (Fig.~\ref{Fig_worldlines}a) that depict virtual pair
creation-annihilation events; in the nematic phase arbitrarily long
dislocation worldlines exist and may entangle
(Fig.~\ref{Fig_worldlines}b), similar to worldlines of particles in a
Bose superfluid~\cite{Feynman_book}.

Now I will present a mathematical formalism supporting
these qualitative considerations. It is analogous to
the duality transformation introduced in Ref.~\cite{Fisher_89}.

%
%

Our first step is to incorporate the dislocations into the effective
action~(\ref{A_sm}). This is accomplished by factorizing the smectic
order parameter, $\Psi = \Psi_0 e^{i q_0 u} \times \Psi_D$, where $u$ is
a regular single-valued function and $\Psi_D$ is the phase factor due to
the dislocations. The derivatives of $u$ in Eq.~(\ref{A_sm}) should now
be replaced by $\partial_\mu u - (i / q_0) \Psi_D^* \partial_\mu \Psi_D
\equiv (\partial_\mu u)_{\rm tot}$, $\mu = \tau, x, y$. Decoupling the
quadratic part of the action with a Hubbard-Stratanovich field
$\sigma_\mu$, I get
\begin{eqnarray}
A &=& A_D + \int\limits_0^\beta d \tau \int d^2 {\bf r}
[-i (\partial_\mu u)_{\rm tot} \sigma_\mu - H_a],
\label{A_Hubbard}
\\
H_a &=& -i \frac{C}{Y} l^2 \sigma_x \partial_y \sigma_\tau
+ \frac{\sigma_x^2}{2 Y}
+ \sigma_y (-2 K \partial_y^2)^{-1} \sigma_y
\nonumber\\
\mbox{} &+& \frac{l^4}{2} \partial_y \sigma_\tau
\left({\bf U} - \frac{C^2}{Y}\right)
\partial_y \sigma_\tau,
\label{H_a}
\end{eqnarray}
where $\hbar = 1$, $\sigma_\tau \equiv p / l^2$, and $A_D$ contains
terms describing dislocation cores (see below). Integration over $u$
gives the constraint $\partial_\mu \sigma_\mu = 0$, which I implement by
means of an auxillary $U(1)$ gauge field $a_\mu$,
\begin{equation}
\sigma_\mu = \epsilon_{\mu\nu\lambda}
\partial_\nu a_\lambda \equiv [\partial \times a]_\mu.
\label{a}
\end{equation}
This leads to the (dual) action of the form
\begin{equation}
A_{\rm dual} = A_D + \int\limits_0^\beta d \tau \int d^2 {\bf r}
 (-i \Lambda a_\mu j_\mu + H_a),
\label{A_dual_first}
\end{equation}
where $\Lambda = 2 \pi / q_0$ and $j_\mu = (2 \pi i)^{-1}
\epsilon_{\mu\nu\lambda} \partial_\nu (\Psi_D^* \partial_\lambda
\Psi_D)$ is the dislocation 3-current. This current can be expressed in
terms of the second-quantized dislocation field $\Phi$,
\begin{equation}
j_\mu = t_\mu \Phi^* (-i \partial_\mu +  \Lambda a_\mu) \Phi,
\label{j_mu}
\end{equation}
which is assumed to be bosonic. Other types of quantum statistics
for dislocations are exotic alternatives, which are
not investigated here.

Making standard  assumptions about $A_D$, I arrive at
\begin{eqnarray}
A_{\rm dual} &=& \int\limits_0^\beta d \tau \int d^2 {\bf r}
 \biggl\{\frac{t_\mu}{2}|(-i \partial_\mu - \Lambda a_\mu
 - e_D a_\mu^{\rm ext})
\Phi|^2\biggr.
\nonumber\\
\mbox{} &+& \biggl.V(\Phi) + H_a[a]\biggr\}.
\label{A_dual}
\end{eqnarray}
The obtained dual theory, Eqs.~(\ref{H_a}), (\ref{a}), and
(\ref{A_dual}) describes $\Phi$ bosons interacting with each other and
with the $U(1)$ gauge field $a_\mu$. The phenomenological parameters
introduced above are as follows. Parameter $t_\tau \sim \hbar^2 / E_c$,
where $E_c$ is the dislocation core energy. It was estimated within the
Hartree-Fock approximation in Refs.~\cite{Fogler_96} and
\cite{Wexler_xxx}. Near $\nu \sim \nu_c$ where the interstripe
separation $\Lambda$ approaches the quantum fluctuation scale $l$, $E_c$
is expected to be much smaller than the Hartree-Fock value, but any
precise estimate is challenging. Parameter $t_x$ of dimension of ${\rm
energy} \times {\rm (length)}^2$ is the hopping matrix element for
dislocation motion in the $\hat{\bf x}$-direction, i.e., dislocation
{\it glide\/}. Such a glide requires quantum tunneling and is
exponentially small unless $\Lambda \lesssim l$. Parameter $t_y$
describes the dislocation climb, which also originates from the dynamics
on the microscopic length scales. One may recall that the climb requires
mobile point defects. Although such defects are not among fundamental
low-energy excitations of the theory, composite point defects, in the
form of short stripe segments or dislocation pairs, do exist. They are
most likely short lived because dislocations are not expected to form
bound pairs in the nematic phase. Yet during their limited life span
such composite defects can perfectly well assist the climb by
micromotions of constituent electrons. In general, I am not aware of any
fundamental principle that would protect $t_y = 0$ value, although it
seems reasonable that $t_y \ll t_x$ in our case.

Yet another phenomenological variable is the potential $V(\Phi) = m_\Phi
|\Phi|^2 + r_\Phi |\Phi|^4 + \ldots$, which accounts for a self-energy
and a short-range interaction between the dislocations; the scales of
$m_\Phi$ and $r_\Phi$ are set by $E_c$ and $E_c \Lambda^2$,
respectively. Finally, $e_D$ is electric charge of the dislocation that
couples to the external vector potential $a_\tau^{\rm ext} = a_x^{\rm
ext} = 0$, $a_y^{\rm ext} = B x$. This coupling is introduced only for
the sake of generality. Since I study electron liquid crystal phases
derived from incompressible liquids, I expect dislocations to be
electrically neutral, i.e., $e_D = 0$.

Another few comments are in order. The derived theory is meant to
capture only the dynamics of neutral (dipolar) excitations of the
system. The underlying incompressible state has its own dynamics
characterized, most importantly, by the quantized Hall conductivity.

There is a very strong similarity between our theory and the statistical
mechanics model of the 3D smectic-nematic transition studied
by Toner and others~\cite{Toner_82}.

Integrating out the gauge field from $A_{\rm dual}$ leads to the model
of dislocation lines interacting via an effective Biot-Savart potential
(which in our case is short-range for Coulombic
$U$~\cite{Fogler_in_preparation}). One may argue~\cite{Helfrich_78} that
the unlimited expansion of the dislocation loops occurs when the energy
cost for creating a large loop of length $L$ is compensated by the
``entropic gain'' $L / l_0$, where the persistence length $l_0$ is
determined by short-range physics. Despite the physical appeal of this
argument, the gauge theory~(\ref{A_dual}) is presumably better suited
for a quantitative analysis (perhaps, along the lines of
Ref.~\cite{Toner_82}).

{\it Phases and their collective modes\/}.--- Let us now see how the
smectic and nematic states are reproduced in the dual theory. As
discussed above, the smectic phase corresponds to $\langle \Phi \rangle
= 0$. In this case $A_{\rm dual}$ reduces to $H_a$, which is quadratic.
The low-energy dynamics is that of a gas of noninteracting Goldstone
bosons, which are the aforementioned magnetophons. It is a simple
exercise to verify that their dispersion relation is given by
Eq.~(\ref{omega_smectic}).

In the nematic phase dislocations have condensed, $\langle \Phi \rangle
= \Phi_0 \neq 0$. In conventional local $U(1)$ gauge theories,
the appearance of such an order parameter triggers the Anderson-Higgs
mechanism, eliminating the gapless Goldstone modes. This can not be the
case here because the nematic state does break the continuous symmetry
with respect to spatial rotations. The seeming paradox is resolved due
to the peculiar feature of the present gauge theory: the nonlocality of
the gauge-field strength term $H_a$, see Eq.~(\ref{H_a}). By virtue of
that, the condensation of $\Phi$ merely stiffens the collective mode,
leaving it gapless. Neglecting terms proportional to $C$, I find that
\begin{equation}
\omega_1({\bf q}) =
 \left(\frac{m_x}{m_\tau} q_x^2 + m_x K q_y^2\right)^{1/2},
\:\: m_\mu \equiv t_\mu \Lambda^2 |\Phi_0|^2.
\label{omega_nematic_1}
\end{equation}
As for the magnetophonon mode of the smectic~(\ref{omega_smectic}), it
indeed acquires a small gap $\sqrt{m_y Y}$ at $q = 0$. It anti-crosses
with the acoustic branch~(\ref{omega_nematic_1}) near the point
$\omega_1^2({\bf q}) \sim m_y Y$, and at larger $q$ becomes the lowest
frequency collective mode with the dispersion relation
\begin{equation}
\omega_2({\bf q}) = \left[\frac{q^2_x q^2_y}{m^2 \omega_c^2} Y U(q)
                  + m_y Y\right]^{1/2}
\label{omega_nematic_2}
\end{equation}
only slightly different from~(\ref{omega_smectic}). At such $q$ the
structure factor of the nematic has two sets of $\delta$-functional
peaks,
%
\[
S(\omega, {\bf q}) = \frac{\pi \hbar q_y^2}{m \omega_c^2}
\left[\frac{K q_y^4}{m n_0} \delta(\omega^2 - \omega_1^2) +
\frac{Y q_x^2}{m n_0} \delta(\omega^2 - \omega_2^2)\right],
\]
%
which split between themselves the spectral weight of the single
collective mode of the smectic. The presence of the two modes can be
explained by the existence of two order parameters: a unit vector (more
precisely, director) ${\bf N}$ normal to the local stripe orientation
and the complex wavefunction $\Phi_0$ of the dislocation condensate.
Classical 2D nematics have two (overdamped) modes virtually for the same
reason~\cite{Toner_81}.

In conclusion, let us address measurable properties of the novel quantum
Hall states considered here. At low temperature both the parent
incompressible state and its liquid crystal descendants will be
insulating. If $T$ is not too small, formation of stripe superstructures
can be deduced from the anisotropic
magnetoresistance~\cite{Experiments}. On the other hand, the microwave
absorption will be anisotropic even at low $T$ and would enable one to
further distinguish between the smectic and nematic phases: the nematic
will show two dispersing collective modes while the smectic will produce
a single one. To circumvent disorder pinning effects, such measurements
should be done at high enough $q$.

{\it Acknowledgements}---This work is supported by MIT Pappalardo
Program in Physics. I wish to thank X.-G. Wen for useful discussions,
and also A.~Dorsey, L.~Radzihovsky, and C.~Wexler for valuable comments
on the manuscript.


\begin{thebibliography}{29}
\bibitem{Comment_on_soft_modes} I am interested in a regime
with very strong quantum fluctuations where the smectic order parameter
is small and only the main harmonic of the soft mode is
important.
\bibitem{Comment_on_compressible} If the parent state is compressible,
the additional low-energy degrees of freedom would generate an effective
viscosity of Landau damping (or Caldeira-Leggett) type entering
Eq.~(\ref{H}). Its consequences are left for future study.
\bibitem{Comment_on_commutator} In principle, $\delta n = -\partial_y p
+ f(u)$ with some function $f$. In the absence of dislocations the most
relevant term is $f = A \partial_x u$, which leads again to
Eq.~(\ref{A_sm}). If dislocations are allowed, the situation is more
complicated, see later in the main text.
\end{thebibliography}
\end{document}